\documentclass[]{article}
\usepackage{latexsym}
\usepackage{epsfig}

\Large

\long\def\@makecaption#1#2{%
  \vskip\abovecaptionskip
  \sbox\@tempboxa{\pushziti\small\rm\songti\zihao{-5}#1: #2\popziti}%
  \ifdim \wd\@tempboxa >\hsize
    \box\@tempboxa\par
  \else
    \global \@minipagefalse
    \hbox to\hsize{\hfil\box\@tempboxa\hfil}%
  \fi
  \vskip\belowcaptionskip}

\begin{document}

\centerline{\Large{\bf Governing Equations of Compressible
                          Turbulence (The Revised)}}

$${}$$
\centerline{Feng \quad Wu }\centerline{\footnote{klfjgjg}}
\centerline{\it Department of Mechanics
and Mechanical Engineering,}

\centerline{\it University of Science and Technology of China, Hefei 230026, China}
$${}$$
\setlength{\baselineskip}{25pt}

\noindent \small{By the nonstandard analysis theory of turbulence,
the governing equations of compressible turbulence are given. The
equations can hold at non-uniform points, in fact, are new kind of
equations. There are three choices. In the choice one, the
second-order infinitesimal quantities are neglected, the closed
equations about the point(monad)-average quantities are obtained.
In other two choices, the closed equations, in which the
third-order infinitesimal quantities are omitted, are given and
about the instantaneous, point-averaged and fluctuant quantities.
}

$${}$$

\large \setlength{\baselineskip}{25pt}

\indent In the paper \cite{lgc}, a new approach, the nonstandard
picture of turbulence, is presented. The essential idea in this
picture is that a particle of fluid in a laminar field is uniform
wholly, but in turbulence a particle of fluid is not uniform and
has interior structure. By the nonstandard analysis, this picture
can be described in mathematics. In the nonstandard analysis
mathematics, an infinitesimal $\varepsilon$ is a certain
number(nonstandard number) rather than a process of tending to
zero. A particle of fluid is called as a monad and the dimension
of a monad is an infinitesimal $\varepsilon$. By the concepts of
the nonstandard analysis, the definition of ``differential" can be
given:
\begin{equation}
\frac{\partial f}{\partial
t}=\frac{f(t+\varepsilon)-f(t)}{\varepsilon},\quad \frac{\partial
f}{\partial x}=\frac{f(x+\varepsilon)-f(x)}{\varepsilon}
\end{equation}
There is conceptual difference between this definition and that of
the differential in the standard analysis theory.

In the nonstandard analysis theory of turbulence(NATT), there are
six assumptions. They are:

\begin{quote}
    \it{Assumption 1: Global turbulent field is composed of standard
   points, and every standard point is yet a monad. Each monad possesses
   the internal structure, namely a monad is also composed of infinite
   nonstandard points (so called interior points of the monad)}.
\end{quote}

\begin{quote}
     \it{Assumption 2: The flows in monad fields are controlled by the
      Navier-Stokes equations.}.
\end{quote}

\begin{quote}
     \it{Assumption 3: Turbulent field is continuous}.
\end{quote}

\begin{quote}
     \it{Assumption 4: When a measurement at any point (monad) $(x_{1},x_{2},x_{3},t)$
     in a physical field is taken, the operation of the measurement
    will act randomly on one interior point (nonstandard point) of the point $(x_{1},x_{2},
    x_{3},t)$}.
\end{quote}

\begin{quote}
     \it{Assumption 5: When a measurement at any point (monad) of a turbulent
    field is made, the operation of the measurement will act in equiprobability on various
    interior points of the monad. This assumption is called the equiprobability
    assumption}.
\end{quote}

\begin{quote}
  \it{Assumption 6: In both the value and structure
  of function, physical function, defined on the interior points of the monads
  of a turbulent field, is very close between two monads,  when these two monads
  are infinitely close to each other}.
\end{quote}

By virtue of these assumptions, the fundamental equations for
incompressible turbulence are obtained, also the closure problem
is overcome in the paper \cite{lgc}. These equations are based on
the definition (1) and new kind of equations, which can hold at
non-uniform points.

Now using the concepts mentioned above, we will give the governing
equations for compressible turbulence.

The equations, the Navier-Stokes equations, governing the motion
of laminar flows hold only at uniform points. The nonstandard
points in a monad are uniform, therefore the Navier-Stokes
equations hold in monad fields.

In a monad field, the governing equations of compressible flows
are as follows.
\begin{equation}
\frac{\partial \rho}{\varepsilon^{2}\partial
t'}+\frac{\partial(\rho U_{i})}{\varepsilon^{2}\partial x'_{i}}=0
\end{equation}
\begin{equation}
\frac{\partial(\rho U_{i})}{\varepsilon^{2}\partial
t'}+\frac{\partial(\rho U_{i}U_{j})}{\varepsilon^{2}\partial
x'_{j}}=-\frac{\partial P}{\varepsilon^{2}\partial
x'_{i}}+\frac{\partial t_{ij}}{\varepsilon^{2}\partial x'_{j}}
\end{equation}
\begin{equation}
\frac{\partial}{\varepsilon^{2}\partial t'}\left
[\rho(e+\frac{1}{2}U_{i}U_{i})\right]+\frac{\partial}{\varepsilon^{2}\partial
x'_{j}}\left[\rho
U_{j}(h+\frac{1}{2}U_{i}U_{i})\right]=\frac{\partial}{\varepsilon^{2}\partial
x'_{j}}(U_{i}t_{ij})-\frac{\partial q_{j}}{\varepsilon^{2}\partial
x'_{j}}
\end{equation}

\noindent Here the rule of summation over repeated indices is
adopted, and $U_{i}$ is the velocity component in $i-$direction,
$P$ the pressure, $\rho$ the density, $e$ the intrinsic energy per
unit mass, $h$ the enthalpy per unit mass. $q_{j}$ is the heat
conduction,
\begin{equation}
h=e+\frac{P}{\rho}
\end{equation}
\begin{equation}
P=f(\rho,T)
\end{equation}
\begin{equation}
q_{j}=-\kappa\frac{\partial T}{\varepsilon^{2}\partial x'_{j}}
\end{equation}

\noindent $t_{ij}$ is the stress tensor, $t_{ji}=t_{ij}$,
\begin{equation}
t_{ij}=\mu\left(\frac{\partial U_{i}}{\varepsilon^{2}\partial
x'_{j}}+\frac{\partial U_{j}}{\varepsilon^{2}\partial
x'_{i}}\right)-\frac{2}{3}\mu\delta_{ij}\frac{\partial
U_{d}}{\varepsilon^{2}\partial x'_{d}}
\end{equation}
$$S_{ij}=\frac{1}{2}\left(\frac{\partial U_{i}}{\varepsilon^{2}\partial
x'_{j}}+\frac{\partial U_{j}}{\varepsilon^{2}\partial
x'_{i}}\right)$$

\noindent Here $\mu$ is the dynamic viscosity, $\kappa$ the
thermal conductivity, $T$ the absolute temperature, $\delta_{ij}$
the Kronecker delta.

The independent variables of all functions in these equations are

\noindent $(x_{1},x_{2},x_{3},t,x'_{1},x'_{2},x'_{3},t')$.
$(x_{1},x_{2},x_{3},t)$ is the coordinates set up in global field,
while $(x'_{1},x'_{2},x'_{3},t')$ is the coordinates set up in a
monad \cite{lgc}.

Still there are the relations:
$$U_{i}=\widetilde{U_{i}}+u_{i},\quad P=\widetilde{P}+p,\quad
\rho=\widetilde{\rho}+\rho_{f},\quad T=\widetilde{T}+T_{f},\quad
e=\widetilde{e}+e_{f}$$
\begin{equation}
h=\widetilde{h}+h_{f},\quad
q_{j}=\widetilde{q_{j}}+(q_{j})_{f},\quad
t_{ij}=\widetilde{t_{ij}}+(t_{ij})_{f}\quad \quad etc.
\end{equation}

\noindent Here the lower-case $u_{i},p$ and the quantities with
index $f$ are called as fluctuation quantities, which have the
order of magnitude $0(\varepsilon$).

Let $``\sim"$ express the average operation over
$(x'_{1},x'_{2},x'_{3},t')$, $``\sim^{i}"$ over $x'_{i}$, $``-"$
over $t'$, etc., in a monad. Therefore,
\begin{equation}
\widetilde{U}=\frac{1}{L_{t}}\int_{0}^{L_{t}}dt'\frac{1}{L^{3}}
\int_{0}^{L}\int_{0}^{L}\int_{0}^{L}Udx'_{1}dx'_{2}dx'_{3}
\end{equation}

\noindent Here $L_{t}$ and $L$ are the infinite of time and space
respectively.

Moreover, by use of the method used in the paper \cite{lgc}, the
following equations are obtained,

\begin{equation}
\frac{\partial\widetilde{\rho}}{\partial
t}+\frac{\partial\widetilde{(\rho U_{i})}}{\partial x_{i}}=0
\end{equation}
\begin{equation}
\frac{\partial\widetilde{(\rho U_{i})}}{\partial
t}+\frac{\partial\widetilde{(\rho U_{i}U_{j})}}{\partial x_{j}}=-
\frac{\partial\widetilde{P}}{\partial
x_{i}}+\frac{\partial\widetilde{t_{ij}}}{\partial x_{j}}
\end{equation}
\begin{equation}
\frac{\partial}{\partial t}
\widetilde{[\rho(e+\frac{1}{2}U_{i}U_{i})]}+\frac{\partial}{\partial
x_{j}}\widetilde{[\rho
U_{j}(h+\frac{1}{2}U_{i}U_{i})]}=\frac{\partial}{\partial
x_{j}}\widetilde{(U_{i}t_{ij})}-\frac{\partial\widetilde{
q_{j}}}{\partial x_{j}}
\end{equation}

The equations (11)-(13), in fact, are conservation equations of
mass, momentum and energy, respectively, at one monad of global
field. By the relations (9), we can obtained, from (11)-(13), the
instantaneous and fluctuant equations. Clearly equation (11) can
be written as
\begin{equation}
\frac{\partial \rho}{\partial t}-\frac{\partial \rho_{f}}{\partial
t}+\frac{\partial(\rho U_{i})}{\partial x_{i}}-\frac{\partial(\rho
U_{i})_{f}}{\partial x_{i}}=0
\end{equation}

\noindent After splitting the equation (14) into two parts in
different order of magnitude, it follows that:

\noindent The fluctuant continuity-equation in the order of
magnitude $0(\varepsilon)$ is:
\begin{equation}
\frac{\partial \rho_{f}}{\partial t}+\frac{\partial(\rho
U_{i})_{f}}{\partial x_{i}}=0
\end{equation}

\noindent The instantaneous continuity-equation in the order of
magnitude $0(1)$ is:
\begin{equation}
\frac{\partial \rho}{\partial t}+\frac{\partial(\rho
U_{i})}{\partial x_{i}}=0
\end{equation}

Similarly, from (12) and (13), we have the instantaneous
momentum-equations and energy-equation,
\begin{equation}
\frac{\partial}{\partial t}(\rho U_{i})+\frac{\partial}{\partial
x_{j}}(\rho U_{i}U_{j})=-\frac{\partial P}{\partial
x_{i}}+\frac{\partial t_{ij}}{\partial x_{j}}
\end{equation}
\begin{equation}
\frac{\partial}{\partial
t}\left[\rho(e+\frac{1}{2}U_{i}U_{i})\right]+\frac{\partial}{\partial
x_{j}}\left[\rho
U_{j}(h+\frac{1}{2}U_{i}U_{i})\right]=\frac{\partial}{\partial
x_{j}}(U_{i}t_{ij})-\frac{\partial q_{j}}{\partial x_{j}}
\end{equation}

\noindent also the fluctuant momentum-equations and
energy-equation,
\begin{equation}
\frac{\partial}{\partial t}(\rho
U_{i})_{f}+\frac{\partial}{\partial x_{j}}(\rho
U_{i}U_{j})_{f}=-\frac{\partial p}{\partial x_{i}}+\frac{\partial
(t_{ij})_{f}}{\partial x_{j}}
\end{equation}
\begin{equation}
\frac{\partial}{\partial
t}\left[\rho(e+\frac{1}{2}U_{i}U_{i})\right]_{f}+\frac{\partial}{\partial
x_{j}}\left[\rho
U_{j}(h+\frac{1}{2}U_{i}U_{i})\right]_{f}=\frac{\partial}{\partial
x_{j}}(U_{i}t_{ij})_{f}-\frac{\partial (q_{j})_{f}}{\partial
x_{j}}
\end{equation}

Now let $P=R\rho T,\quad e=c_{v}T$ (in the case of perfect gas).
Here $R$ and $ c_{v}$ are, respectively, the gas constant and the
specific heat at constant volume. So there are the relations as
follows.
\begin{equation}
h=e+\frac{P}{\rho}=e+\frac{\widetilde{P}}{\widetilde{\rho}}
\left[1+\frac{p}{\widetilde{P}}-\frac{\rho_{f}}{\widetilde{\rho}}
-\frac{p\rho_{f}}{\widetilde{P}\widetilde{\rho}}
+\frac{\rho_{f}\rho_{f}}{\widetilde{\rho}\widetilde{\rho}}\right]+0(\varepsilon^{3})
\end{equation}
\begin{equation}
\widetilde{h}=\widetilde{e}+\frac{\widetilde{P}}{\widetilde{\rho}}
\left[1-\frac{\widetilde{(p\rho_{f})}}{\widetilde{P}\widetilde{\rho}}
+\frac{\widetilde{(\rho_{f}\rho_{f})}}{\widetilde{\rho}\widetilde{\rho}}\right]+0(\varepsilon^{3})
\end{equation}
\begin{equation}
h_{f}=h-\widetilde{h}=e_{f}+\frac{\widetilde{P}}{\widetilde{\rho}}
\left[\frac{p}{\widetilde{P}}-\frac{\rho_{f}}{\widetilde{\rho}}
-\frac{p\rho_{f}-\widetilde{(p\rho_{f})}}{\widetilde{P}\widetilde{\rho}}
+\frac{\rho_{f}\rho_{f}-\widetilde{(\rho_{f}\rho_{f})}}
{\widetilde{\rho}\widetilde{\rho}}\right]+0(\varepsilon^{3})
\end{equation}
\begin{equation}
P=R\rho T=R[\widetilde{\rho}\widetilde{T}+\widetilde{\rho}T_{f}
+\rho_{f}\widetilde{T}+\rho_{f}T_{f}]
\end{equation}
\begin{equation}
\widetilde{P}=R[\widetilde{\rho}\widetilde{T}+\widetilde{(\rho_{f}
T_{f})}]
\end{equation}
\begin{equation}
p=P-\widetilde{P}=R[\widetilde{\rho}T_{f}+\rho_{f}\widetilde{T}
+\rho_{f}T_{f}-\widetilde{(\rho_{f}T_{f})}]
\end{equation}

Yet by Assumption 6, $\frac{\partial}{\partial
x_{j}}[u_{i}u_{j}-\widetilde{(u_{i}u_{j})}]\sim
0(\varepsilon^{3})$. And so are the similar others. When the terms
in order of magnitude $\sim 0(\varepsilon^{3})$ are, proximately,
neglected, those will be omitted.

Then the expansion of $\rho U_{i}$, $\widetilde{(\rho U_{i})}$ and
$(\rho U_{i})_{f}$ is, for example, given as follows. And so are
the similar others.
$$\rho U_{i}=(\widetilde{\rho}+\rho_{f})(\widetilde{U_{i}}+u_{i})=
\widetilde{\rho}\widetilde{U_{i}}+\rho_{f}\widetilde{U_{i}}+
\widetilde{\rho}u_{i}+\rho_{f}u_{i}$$
$$\widetilde{(\rho U_{i})}=\widetilde{\rho}\widetilde{U_{i}}+\widetilde{\rho_{f} u_{i}}$$
\begin{equation}
(\rho U_{i})_{f}=\rho U_{i}-\widetilde{(\rho
U_{i})}=\widetilde{\rho}u_{i}+\rho_{f}\widetilde{U_{i}}
+\rho_{f}u_{i}-\widetilde{(\rho_{f}u_{i})}
\end{equation}

Now by the Assumption 6 and using the expansion like (27), from
(11)-(13), we can write the mean equations:
\begin{equation}
\frac{\partial \widetilde{\rho}}{\partial
t}+\frac{\partial}{\partial
x_{i}}(\widetilde{\rho}\widetilde{U_{i}})+\frac{\partial}{\partial
x_{i}}(\rho_{f}u_{i})+0(\varepsilon^{3})=0
\end{equation}

$$\frac{\partial}{\partial
t}(\widetilde{\rho}\widetilde{U_{i}})+\frac{\partial}{\partial
x_{j}}(\widetilde{\rho}\widetilde{U_{i}}\widetilde{U_{j}})+
\frac{\partial}{\partial
t}(\rho_{f}u_{i})+\frac{\partial}{\partial
x_{j}}(\widetilde{\rho}u_{i}u_{j}+\rho_{f}\widetilde{U_{i}}u_{j}
+\rho_{f}u_{i}\widetilde{U_{j}})$$
\begin{equation}=-\frac{\partial\widetilde{P}}{\partial
x_{i}}+\frac{\partial\widetilde{(t_{ij})}}{\partial
x_{j}}+0(\varepsilon^{3})
\end{equation}

$$\frac{\partial}{\partial t}\left[\widetilde{\rho}\widetilde{e}
+\frac{1}{2}\widetilde{\rho}\widetilde{U_{i}}\widetilde{U_{i}}+\rho_{f}e_{f}+
\frac{1}{2}\widetilde{\rho}u_{i}u_{i}+\rho_{f}\widetilde{U_{i}}u_{i}\right]+$$
$$\frac{\partial}{\partial x_{j}}\left[\widetilde{\rho}\widetilde{U_{j}}
(\widetilde{h}+\frac{\widetilde{P}}{\widetilde{\rho}}
\frac{\rho_{f}\rho_{f}}{\widetilde{\rho}\widetilde{\rho}}
-\frac{p\rho_{f}}{\widetilde{\rho}\widetilde{\rho}})+
\widetilde{\rho}u_{j}h_{f}+\rho_{f}\widetilde{U_{j}}h_{f}+\rho_{f}u_{j}\widetilde{h}\right]+$$
$$\frac{\partial}{\partial x_{j}}\left[\frac{1}{2}\widetilde{\rho}\widetilde{U_{j}}
\widetilde{U_{i}}\widetilde{U_{i}}+\frac{1}{2}\widetilde{\rho}\widetilde{U_{j}}u_{i}u_{i}+
\frac{1}{2}\rho_{f}u_{j}\widetilde{U_{i}}\widetilde{U_{i}}+\widetilde{\rho}u_{j}\widetilde{U_{i}}u_{i}+
\rho_{f}\widetilde{U_{j}}\widetilde{U_{i}}u_{i}\right]$$
\begin{equation}
=\frac{\partial}{\partial
x_{j}}\left[\widetilde{U_{i}}\widetilde{t_{ij}}+u_{i}(t_{ij})_{f}\right]-
\frac{\partial\widetilde{q_{j}}}{\partial
x_{j}}+0(\varepsilon^{3})
\end{equation}

\noindent and, from (15),(19) and (20),  the fluctuant equations:
\begin{equation}
\frac{\partial \rho_{f}}{\partial t}+\frac{\partial}{\partial
x_{i}}(\widetilde{\rho}u_{i})+\frac{\partial}{\partial
x_{i}}(\rho_{f}\widetilde{U_{i}})+0(\varepsilon^{3})=0
\end{equation}

$$\frac{\partial}{\partial
t}(\widetilde{\rho}u_{i})+\frac{\partial}{\partial
t}(\rho_{f}\widetilde{U_{i}})+\frac{\partial}{\partial
x_{j}}(\widetilde{\rho}\widetilde{U_{i}}u_{j}+
\widetilde{\rho}u_{i}\widetilde{U_{j}}+
\rho_{f}\widetilde{U_{i}}\widetilde{U_{j}})$$
\begin{equation}=-\frac{\partial p}{\partial
x_{i}}+\frac{\partial(t_{ij})_{f}}{\partial
x_{j}}+0(\varepsilon^{3})
\end{equation}

$$\frac{\partial}{\partial t}\left[\widetilde{\rho}e_{f}
+\rho_{f}\widetilde{e}+\frac{1}{2}\rho_{f}\widetilde{U_{i}}\widetilde{U_{i}}+
\widetilde{\rho}\widetilde{U_{i}}u_{i}\right]+$$
$$\frac{\partial}{\partial x_{j}}\left[\widetilde{\rho}\widetilde{U_{j}}h_{f}+
\widetilde{\rho}u_{j}\widetilde{h}+
\rho_{f}\widetilde{U_{j}}\widetilde{h}\right]+
\frac{\partial}{\partial
x_{j}}\left[\frac{1}{2}\widetilde{\rho}u_{j}
\widetilde{U_{i}}\widetilde{U_{i}}+\frac{1}{2}\rho_{f}\widetilde{U_{j}}
\widetilde{U_{i}}\widetilde{U_{i}}+
\widetilde{\rho}\widetilde{U_{j}}\widetilde{U_{i}}u_{i}\right]$$
\begin{equation}
=\frac{\partial}{\partial
x_{j}}\left[\widetilde{U_{i}}(t_{ij})_{f}+u_{i}\widetilde{t_{ij}}\right]-
\frac{\partial(q_{j})_{f}}{\partial x_{j}}+0(\varepsilon^{3})
\end{equation}

Finally, from the equations (16)-(18), (28)-(30) and (31)-(33),
the closed equations of turbulence in compressible fluid can be
easily obtained. There are, like the case of incompressible
turbulence, three choices.

Choice one: In the equations (28)-(30), the terms in the order of
magnitude $0(\varepsilon^{2})$ are omitted. We have
\begin{equation}
\frac{\partial\widetilde{\rho}}{\partial
t}+\frac{\partial}{\partial
x_{i}}(\widetilde{\rho}\widetilde{U_{i}})=0
\end{equation}
\begin{equation}
\frac{\partial}{\partial t}(\widetilde{\rho}\widetilde{U_{i}})+
\frac{\partial}{\partial
x_{j}}(\widetilde{\rho}\widetilde{U_{i}}\widetilde{U_{j}})=-\frac{\partial\widetilde{P}}{\partial
x_{i}}+\frac{\partial\widetilde{t_{ij}}}{\partial x_{j}}
\end{equation}
\begin{equation}
\frac{\partial}{\partial
t}\left[\widetilde{\rho}(\widetilde{e}+\frac{1}{2}\widetilde{U_{i}}\widetilde{U_{i}})\right]
+\frac{\partial}{\partial
x_{j}}\left[\widetilde{\rho}\widetilde{U_{j}}(\widetilde{h}+
\frac{1}{2}\widetilde{U_{i}}\widetilde{U_{i}})\right]=\frac{\partial}{\partial
x_{j}}(\widetilde{U_{i}}\widetilde{t_{ij}})-\frac{\partial\widetilde{q_{j}}}{\partial
 x_{j}}
\end{equation}

$$\widetilde{P}=R\widetilde{\rho}\widetilde{T},\quad \widetilde{h}=
\widetilde{e}+\frac{\widetilde{P}}{\widetilde{\rho}},\quad
\widetilde{e}=c_{v}\widetilde{T},\quad
\widetilde{q_{j}}=-\kappa\frac{\partial\widetilde{T}}{\partial
x_{j}}$$
\begin{equation}
\widetilde{t_{ij}}=\mu\left(\frac{\partial\widetilde{U_{i}}}{\partial
x_{j}} +\frac{\partial\widetilde{U_{j}}}{\partial
x_{i}}\right)-\frac{2}{3}\mu\delta_{ij}\frac{\partial\widetilde{U_{d}}}{\partial
x_{d}}
\end{equation}
$$$$

Choice two: In the equations (31)-(33), the mean quantities are
written as the differences between instantaneous and fluctuant
quantities. Then the terms in the order of magnitude
$0(\varepsilon^{3})$ are omitted. It follows that

\noindent The instantaneous equations
\begin{equation}
\frac{\partial \rho}{\partial t}+\frac{\partial(\rho
U_{i})}{\partial x_{i}}=0
\end{equation}
\begin{equation}
\frac{\partial}{\partial t}(\rho U_{i})+\frac{\partial}{\partial
x_{j}}(\rho U_{i}U_{j})=-\frac{\partial P}{\partial
x_{i}}+\frac{\partial t_{ij}}{\partial x_{j}}
\end{equation}
\begin{equation}
\frac{\partial}{\partial
t}\left[\rho(e+\frac{1}{2}U_{i}U_{i})\right]+\frac{\partial}{\partial
x_{j}}\left[\rho
U_{j}(h+\frac{1}{2}U_{i}U_{i})\right]=\frac{\partial}{\partial
x_{j}}(U_{i}t_{ij})-\frac{\partial q_{j}}{\partial x_{j}}
\end{equation}

\noindent and the fluctuant equations
\begin{equation}
\frac{\partial \rho_{f}}{\partial t}+\frac{\partial}{\partial
x_{i}}(\rho_{f}U_{i})+\frac{\partial}{\partial x_{i}}(\rho
u_{i})-2\frac{\partial}{\partial x_{i}}(\rho_{f}u_{i})=0
\end{equation}

$$\frac{\partial}{\partial t}(\rho u_{i}+\rho_{f}U_{i}-2\rho_{f}u_{i})
+\frac{\partial}{\partial x_{j}}(\rho U_{i}u_{j}+ \rho u_{i}U_{j}+
\rho_{f}U_{i}U_{j})$$
$$-2\frac{\partial}{\partial x_{j}}(\rho u_{i}u_{j}+\rho_{f}U_{i}u_{j}
+\rho_{f}u_{i}U_{j})$$
\begin{equation}=-\frac{\partial p}{\partial
x_{i}}+\frac{\partial(t_{ij})_{f}}{\partial x_{j}}
\end{equation}

$$\frac{\partial}{\partial t}\left[\rho e_{f}
+\rho_{f}e-2\rho_{f}e_{f}+\frac{1}{2}\rho_{f}U_{i}U_{i}+ \rho
U_{i}u_{i}-\rho u_{i}u_{i}-2\rho_{f}U_{i}u_{i}\right]+$$
$$\frac{\partial}{\partial x_{j}}\left[(\rho u_{j}+\rho_{f}U_{j})h+\rho U_{j}h_{f}
-2(\rho u_{j}+\rho_{f}U_{j})h_{f}-2\rho_{f}u_{j}h+2\rho
U_{j}\left(\frac{p\rho_{f}}{\rho\rho}-\frac{P\rho_{f}\rho_{f}}{\rho\rho\rho}\right)\right]+$$
$$ \frac{\partial}{\partial
x_{j}}\left[\frac{1}{2}\rho u_{j}
U_{i}U_{i}+\frac{1}{2}\rho_{f}U_{j}U_{i}U_{i}+ \rho
U_{j}U_{i}u_{i}-\rho U_{j}u_{i}u_{i}-2\rho
u_{j}U_{i}u_{i}-2\rho_{f}U_{j}U_{i}u_{i}-\rho_{f}u_{j}U_{i}U_{i}\right]$$
\begin{equation}
=\frac{\partial}{\partial
x_{j}}\left[U_{i}(t_{ij})_{f}+u_{i}t_{ij}-2u_{i}(t_{ij})_{f}\right]-
\frac{\partial(q_{j})_{f}}{\partial x_{j}}
\end{equation}

\noindent Here,
$$P=R\rho T,\quad \frac{\partial p}{\partial
x_{i}}=R\frac{\partial}{\partial x_{i}}(\rho
T_{f}+\rho_{f}T-2\rho_{f}T_{f})$$
$$ h=e+\frac{P}{\rho},\quad
h_{f}=e_{f}+\frac{p}{\rho}-\frac{P\rho_{f}}{\rho\rho},\quad
q_{j}=-\kappa\frac{\partial T}{\partial x_{j}},\quad
(q_{j})_{f}=-\kappa\frac{\partial T_{f}}{\partial x_{j}}$$
$$e=c_{v}T,\quad e_{f}=c_{v}T_{f},\quad \quad t_{ij}=\mu\left(\frac{\partial U_{i}}{\partial x_{j}}+
\frac{\partial U_{j}}{\partial
x_{i}}\right)-\frac{2}{3}\mu\delta_{ij}\frac{\partial
U_{d}}{\partial x_{d}}$$
\begin{equation}
(t_{ij})_{f}=\mu\left(\frac{\partial u_{i}}{\partial
x_{j}}+\frac{\partial u_{j}}{\partial
x_{i}}\right)-\frac{2}{3}\mu\delta_{ij}\frac{\partial
u_{d}}{\partial x_{d}}
\end{equation}
$$$$

Choice three: In the equations (28)-(30) and (31)-(33), the terms
in the order of magnitude $0(\varepsilon^{3})$ are omitted. It is
obtained that

\noindent The mean equations
\begin{equation}
\frac{\partial \widetilde{\rho}}{\partial
t}+\frac{\partial}{\partial
x_{i}}(\widetilde{\rho}\widetilde{U_{i}})+\frac{\partial}{\partial
x_{i}}(\rho_{f}u_{i})=0
\end{equation}

$$\frac{\partial}{\partial
t}(\widetilde{\rho}\widetilde{U_{i}})+\frac{\partial}{\partial
x_{j}}(\widetilde{\rho}\widetilde{U_{i}}\widetilde{U_{j}})+
\frac{\partial}{\partial
t}(\rho_{f}u_{i})+\frac{\partial}{\partial
x_{j}}(\widetilde{\rho}u_{i}u_{j}+\rho_{f}\widetilde{U_{i}}u_{j}
+\rho_{f}u_{i}\widetilde{U_{j}})$$
\begin{equation}=-\frac{\partial\widetilde{P}}{\partial
x_{i}}+\frac{\partial\widetilde{(t_{ij})}}{\partial x_{j}}
\end{equation}

$$\frac{\partial}{\partial t}\left[\widetilde{\rho}\widetilde{e}
+\frac{1}{2}\widetilde{\rho}\widetilde{U_{i}}\widetilde{U_{i}}+\rho_{f}e_{f}+
\frac{1}{2}\widetilde{\rho}u_{i}u_{i}+\rho_{f}\widetilde{U_{i}}u_{i}\right]+$$
$$\frac{\partial}{\partial x_{j}}\left[\widetilde{\rho}\widetilde{U_{j}}
(\widetilde{h}+\frac{\widetilde{P}}{\widetilde{\rho}}
\frac{\rho_{f}\rho_{f}}{\widetilde{\rho}\widetilde{\rho}}
-\frac{p\rho_{f}}{\widetilde{\rho}\widetilde{\rho}})+
\widetilde{\rho}u_{j}h_{f}+\rho_{f}\widetilde{U_{j}}h_{f}+\rho_{f}u_{j}\widetilde{h}\right]+$$
$$\frac{\partial}{\partial x_{j}}\left[\frac{1}{2}\widetilde{\rho}\widetilde{U_{j}}
\widetilde{U_{i}}\widetilde{U_{i}}+\frac{1}{2}\widetilde{\rho}\widetilde{U_{j}}u_{i}u_{i}+
\frac{1}{2}\rho_{f}u_{j}\widetilde{U_{i}}\widetilde{U_{i}}+\widetilde{\rho}u_{j}\widetilde{U_{i}}u_{i}+
\rho_{f}\widetilde{U_{j}}\widetilde{U_{i}}u_{i}\right]$$
\begin{equation}
=\frac{\partial}{\partial
x_{j}}\left[\widetilde{U_{i}}\widetilde{t_{ij}}+u_{i}(t_{ij})_{f}\right]-
\frac{\partial\widetilde{q_{j}}}{\partial x_{j}}
\end{equation}

\noindent and fluctuant equations
\begin{equation}
\frac{\partial \rho_{f}}{\partial t}+\frac{\partial}{\partial
x_{i}}(\widetilde{\rho}u_{i})+\frac{\partial}{\partial
x_{i}}(\rho_{f}\widetilde{U_{i}})=0
\end{equation}

$$\frac{\partial}{\partial
t}(\widetilde{\rho}u_{i})+\frac{\partial}{\partial
t}(\rho_{f}\widetilde{U_{i}})+\frac{\partial}{\partial
x_{j}}(\widetilde{\rho}\widetilde{U_{i}}u_{j}+
\widetilde{\rho}u_{i}\widetilde{U_{j}}+
\rho_{f}\widetilde{U_{i}}\widetilde{U_{j}})$$
\begin{equation}=-\frac{\partial p}{\partial
x_{i}}+\frac{\partial(t_{ij})_{f}}{\partial x_{j}}
\end{equation}

$$\frac{\partial}{\partial t}\left[\widetilde{\rho}e_{f}
+\rho_{f}\widetilde{e}+\frac{1}{2}\rho_{f}\widetilde{U_{i}}\widetilde{U_{i}}+
\widetilde{\rho}\widetilde{U_{i}}u_{i}\right]+$$
$$\frac{\partial}{\partial x_{j}}\left[\widetilde{\rho}\widetilde{U_{j}}h_{f}+
\widetilde{\rho}u_{j}\widetilde{h}+
\rho_{f}\widetilde{U_{j}}\widetilde{h}\right]+
\frac{\partial}{\partial
x_{j}}\left[\frac{1}{2}\widetilde{\rho}u_{j}
\widetilde{U_{i}}\widetilde{U_{i}}+\frac{1}{2}\rho_{f}\widetilde{U_{j}}
\widetilde{U_{i}}\widetilde{U_{i}}+
\widetilde{\rho}\widetilde{U_{j}}\widetilde{U_{i}}u_{i}\right]$$
\begin{equation}
=\frac{\partial}{\partial
x_{j}}\left[\widetilde{U_{i}}(t_{ij})_{f}+u_{i}\widetilde{t_{ij}}\right]-
\frac{\partial(q_{j})_{f}}{\partial x_{j}}
\end{equation}

\noindent Here,
$$\frac{\partial\widetilde{P}}{\partial x_{i}}=
R\frac{\partial}{\partial
x_{i}}[\widetilde{\rho}\widetilde{T}+\rho_{f}T_{f}],\quad
\frac{\partial p}{\partial x_{i}}=R\frac{\partial}{\partial
x_{i}}[\widetilde{\rho}T_{f}+\rho_{f}\widetilde{T}]$$
$$\widetilde{h}=\widetilde{e}+
\frac{\widetilde{P}}{\widetilde{\rho}},\quad
h_{f}=e_{f}+\frac{p}{\widetilde{\rho}}
-\frac{\widetilde{P}\rho_{f}}{\widetilde{\rho}\widetilde{\rho}}$$
$$
\widetilde{e}=c_{v}\widetilde{T},\quad e_{f}=c_{v}T_{f},\quad
\widetilde{q_{j}}=-\kappa\frac{\partial\widetilde{T}}{\partial
x_{j}},\quad (q_{j})_{f}=-\kappa\frac{\partial T_{f}}{\partial
x_{j}}$$
$$\widetilde{t_{ij}}=\mu\left(\frac{\partial\widetilde{U_{i}}}{\partial
x_{j}} +\frac{\partial\widetilde{U_{j}}}{\partial
x_{i}}\right)-\frac{2}{3}\mu\delta_{ij}\frac{\partial\widetilde{U_{d}}}{\partial
x_{d}}$$
\begin{equation}
(t_{ij})_{f}=\mu\left(\frac{\partial u_{i}}{\partial
x_{j}}+\frac{\partial u_{j}}{\partial
x_{i}}\right)-\frac{2}{3}\mu\delta_{ij}\frac{\partial
u_{d}}{\partial x_{d}}
\end{equation}
$$$$

We should note that these equations in the three choices are new
kind of equations, which are based on the definition (1) and can
hold at non-uniform points too.

Obviously, the number of unknown quantities equals to that of
equations, therefore the equations in every Choice are closed. The
instantaneous and fluctuant quantities are defined on nonstandard
points $(x_{1},x_{2},x_{3},t,x'_{1},x'_{2},x'_{3},t')$. However,
the mean quantities are the point(monad)-average values. Yet the
average of the point-average values over certain time period or
space range is once again taken. And the results of the again
average could be compared with the measuring mean values over
corresponding time period or space range.

  $${}$$

\end{document}